\documentclass[11pt]{article}
\usepackage{moriond,epsfig}

\def\Journal#1#2#3#4{{#1} {\bf #2}, #3 (#4)}

\def\PRD{{\em Phys. Rev.} D}

\def\MNRAS{\em Mon. Not. R. Astron. Soc.}
\def\ApJ{\em Astrophys. J.}

\def\AA{\em Astron. Astrophys.}

\def\lsim{~\rlap{$<$}{\lower 1.0ex\hbox{$\sim$}}}
\def\bsim{~\rlap{$>$}{\lower 1.0ex\hbox{$\sim$}}}
\def\hmpc{\ {\rm {\it h}^{-1}Mpc}}
\def\hmsun{\ {\rm M_\odot/{\it h}}}

\def\mathbi#1{\textbf{\em #1}}

\def\dc{\delta_{\rm c}}
\def\vk{\mathbi{k}}
\def\vx{\mathbi{x}}
\def\vv{\mathbi{v}}

\def\npk{n_{\rm pk}}

\def\vvpk{\mathbi{v}_{\rm pk}}
\def\xpk{\xi_{\rm pk}}
\def\ppk{P_{\rm pk}}

\def\be{\begin{equation}}
\def\ee{\end{equation}}
\def\bea{\begin{eqnarray}}
\def\eea{\end{eqnarray}}
\begin{document}
\vspace*{4cm}
\title{THE LARGE-SCALE CLUSTERING OF MASSIVE DARK MATTER HALOES}

\author{V. DESJACQUES}

\address{Institute for Theoretical Physics, University of Z\"urich \\
Winterthurerstrasse 190, 8057 Z\"urich, Switzerland}

\maketitle\abstracts{
The statistics of peaks of the initial, Gaussian density field  can be
used to interpret the abundance and clustering of massive dark matter
haloes. I discuss some recent theoretical results related to their
clustering and its redshift evolution. Predictions from the peak model
are qualitatively consistent with measurements of the linear bias  of
high mass haloes, which also show some evidence for a dependence on
the halo mass $M$ at fixed peak height $\nu$. The peak approach also
predicts a distinctive scale-dependence in the bias of haloes across
the baryon acoustic feature, a measurement of which would provide
strong support for its validity. For 2$\sigma$ density peaks
collapsing at $z=0.3$, this residual scale-dependent bias is at the
5-10\% level and should thus be within reach of very large simulations
of structure formation.
}

\section{Peaks in Gaussian random field}

The peak model introduced by \cite{BBKS} assumes that dark matter
haloes are associated with peaks of the initial (Gaussian)  density
field. Although dark matter haloes are the local maxima of the evolved
mass distribution, there is a clear correspondence  with initial
density maxima for massive objects only. In the following, I will
focus on the large-scale clustering properties of initial density
peaks and show there is nontrivial scale-dependence both in the linear
spatial and velocity bias. I will discuss some implications of these
results.

\section{First order biasing of initial density peaks}

Following \cite{BBKS}, one usually smoothes the initial density
fluctuations at redshift $z_i\gg 1$ with a filter of characteristic
mass scale $M$ before identifying local maxima of height $\nu$. Even
though density peaks form a well-behaved point process, the
large-scale asymptotics  of the 2-point correlation and pairwise
velocity can be though of as  arising from the continuous bias
relation \cite{D08,DS10}
\bea
\delta\npk(\vx) &=& b_\nu \delta_M(\vx) - b_\zeta \Delta\delta_M(\vx),
\\ \vvpk(\vx) &=&
\vv_M(\vx)-\frac{\sigma_0^2}{\sigma_1^2}\nabla\delta_M(\vx)\;,
\label{eq:pkbiasing}
\eea 
where $\delta\npk$ and $\vvpk$ are the peak count-in-cell density and
velocity,  $\delta_M$ and $\vv_M$ are the initial mass density and
velocity field smoothed on scale $M$, and the (Lagrangian) bias
parameters $b_\nu$ and $b_\zeta$ are
\be 
b_\nu(\nu,\gamma_1) =
\frac{1}{\sigma_0}\left(\frac{\nu-\gamma_1\bar{u}}{1-\gamma_1^2}\right),
~~~ b_\zeta(\nu,\gamma_1) =
\frac{1}{\sigma_2}\left(\frac{\bar{u}-\gamma_1\nu}{1-\gamma_1^2}\right)\;.
\label{eq:biases}
\ee  
Here, $\bar u\equiv {\bar u}(\nu)$ denotes the mean curvature of peaks
of height $\nu$, $\gamma_1(M)=\sigma_1^2/\sigma_0\sigma_2$ and
$\sigma_0$, $\sigma_1$ and $\sigma_2$ are spectral moments which
depend upon the shape of the linear mass power spectrum. Note that
$b_\zeta$ is strictly positive, whereas $b_\nu$ can be positive or
negative.  In Fourier space, wavemodes of the peak number density
$\delta\npk(\vk)$ can be obtained by multiplying $\delta_M(\vk)$ with
(here and henceforth, I will omit the dependence on $\nu$ and
$\gamma_1$ for brevity)
\be
 b_{\rm pk}(k) = b_\nu + b_\zeta k^2 \;.
\ee
This defines the spatial peak bias at the first order. In practice,
the peak-background split approach, which is based on count-in-cells
statistics, can also be used to estimate $b_\nu$ \cite{CK89}. In this
regards, the linear Lagrangian bias $b_\nu$ predicted by the peak
model is exactly the same as that returned by the peak-background
split argument \cite{DS10}.

The peak velocity $\vvpk(\vx)$ as defined in Eq.(\ref{eq:pkbiasing})
is consistent with the assumption that initial density peaks move
locally with the dark matter.  However, the 3-dimensional velocity
dispersion of peaks is smaller than that of the mass $\sigma_{-1}$,
i.e.  $\sigma^2_{\rm vpk} = \sigma_{-1}^2\, (1 - \gamma_0^2)$ with
$\gamma_0=\sigma_0^2/\sigma_{-1}\sigma_1$, because large-scale flows
are more likely to be directed towards peaks  than to be oriented
randomly \cite{BBKS}. As shown in \cite{DS10}, this leads to a
$k$-dependence of the peak velocities as can been seen upon taking the
divergence of $\vvpk(\vx)$ and Fourier transforming it,
\be
\theta_{\rm pk}(\vk)=\left(1 - \frac{\sigma_0^2}{\sigma_1^2}\,
k^2\right)W\!(k,M)\,\theta(\vk)\equiv b_{\rm vel}(k)\,
\theta_M(\vk)\;,
\label{eq:vpkbiask}
\ee 
where $\theta\equiv\nabla\cdot\vv$ is  the mass velocity divergence
and $W\!(k,M)$ is the Fourier transform of the filter. This defines the
{\it statistical} velocity bias $b_{\rm vel}(k)$. Note that $b_{\rm
vel}(k)$ does not depend on $\nu$ and, for the highest peaks, remains
scale-dependent even though the spatial bias $b_{\rm pk}(k)$ has no
$k$-dependence in this limit.

\section{Redshift evolution of the peak correlation}

Pairwise motions induced by gravitational instabilities will distort
the primeval peak correlation. The gravitational evolution of the
correlation of initial density peaks can be addressed with the
Zel'dovich ansatz \cite{Z70}, assuming they behave like test particles
moving with the dark matter. In this first order approximation, the
gravitationally-evolved peak correlation $\xpk(r,z)$ is the Fourier
transform of the peak power spectrum \cite{D10}
\be
\label{eq:ppk}
\ppk(k,z)=G^2(k,z)\left[b_{\rm vel}(k)+b_{\rm pk}(k,z)\right]^2
P_M(k,0)\;, 
\ee 
where $b_{\rm pk}(k,z)=D(z_i)/D(z)b_{\rm pk}(k)$ and the function 
\be
G^2(k,z)=\left(\frac{D(z)}{D(0)}\right)^2
e^{-\frac{1}{3}k^2\sigma_{\rm vpk}^2\!(z)} 
\ee 
is a damping term induced by velocity diffusion. It is similar to the
propagator $G_\delta(k,z)$ introduced in \cite{CS06}, although  the
latter involves the matter velocity dispersion $\sigma_{-1}$. The
first term in the square bracket reflects the fact that peaks stream
towards (or move apart from) each other in high (low) density
environments, but this effect is $k$-dependent owing to the
statistical velocity bias. Therefore, the Eulerian and Lagrangian
linear bias parameters are related according to
\be
b_\nu^{\rm E}(z)\equiv 1+\frac{D(z_i)}{D(z)}b_\nu(z_i),~~~
b_\zeta^{\rm E}(z)\equiv \frac{D(z_i)}{D(z)}b_\zeta(z_i)
-\frac{\sigma_0^2}{\sigma_1^2}\;.
\ee
The first relation is the usual formula for the Eulerian, linear
scale-independent bias \cite{MW96}. The second relation shows  that
$b_\zeta^{\rm E}$ approaches $-\sigma_0^2/\sigma_1^2$  with time.

\begin{figure}
\centering
\psfig{figure=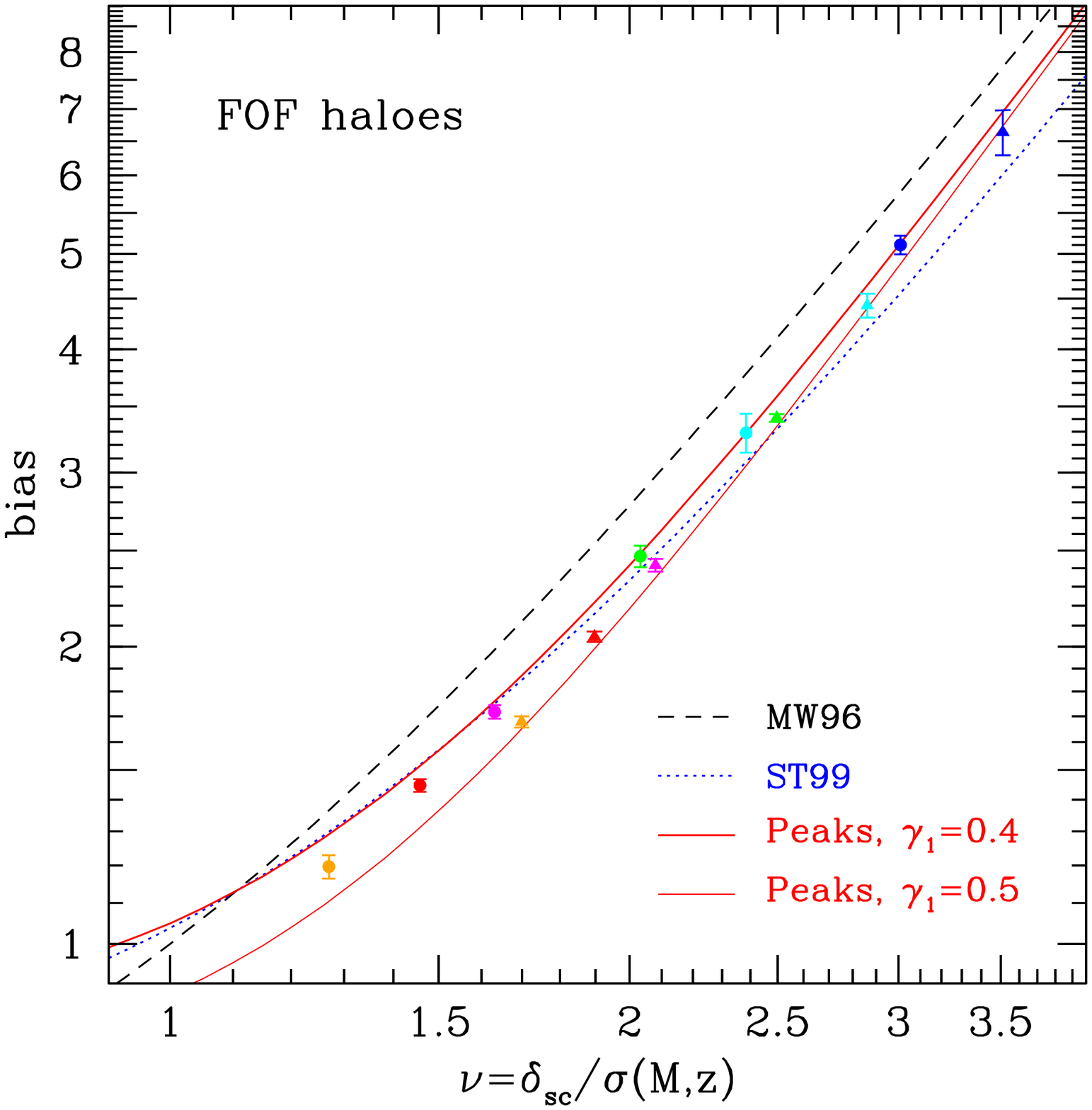,height=2.5in}
\psfig{figure=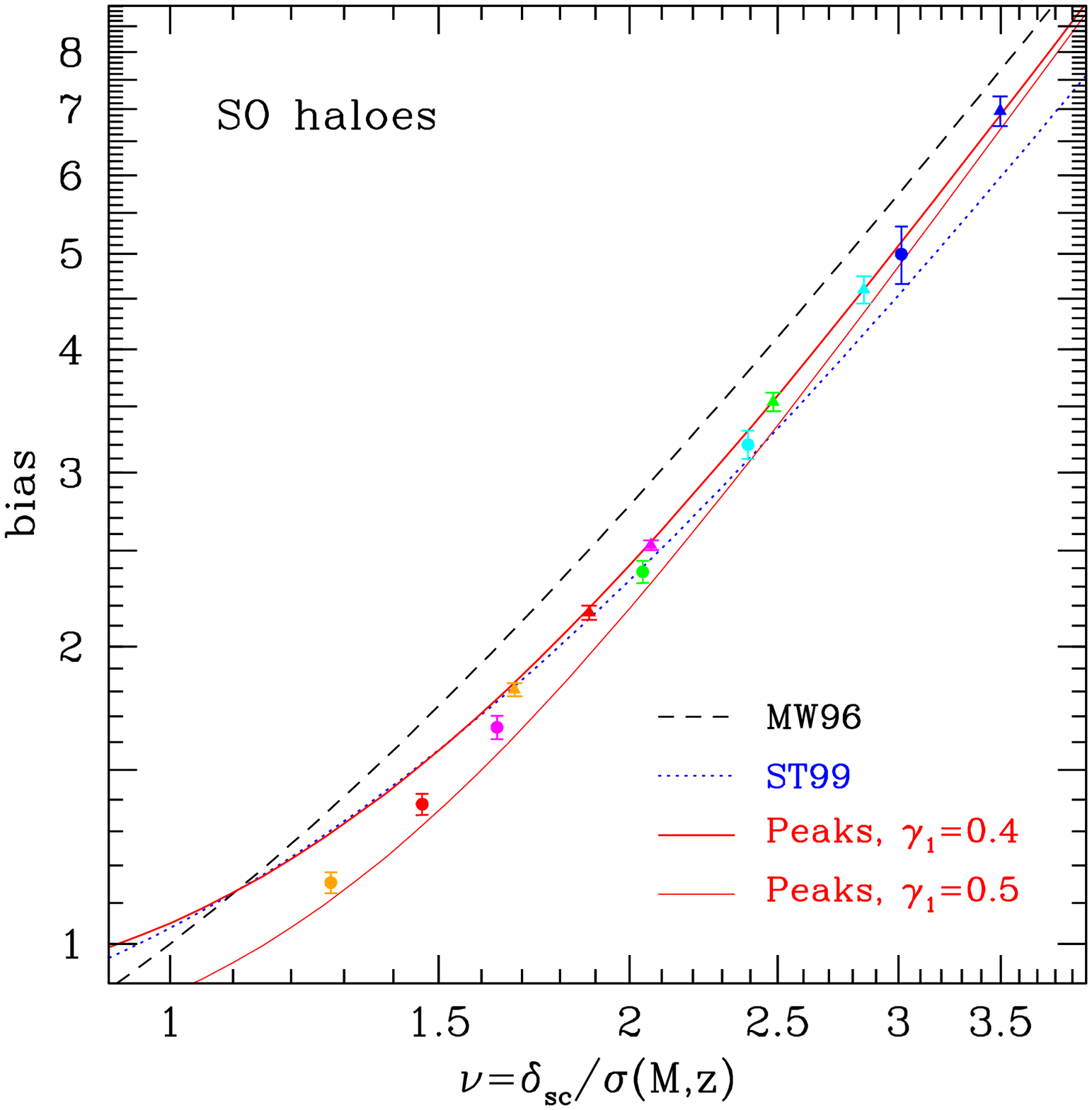,height=2.5in}
\caption{ 
Large-scale bias of dark matter haloes identified with a FOF finder
of linking length $b=0.2$ (left panel) and a SO finder with
redshift-dependent overdensity threshold (right panel).  Circles and
triangles refer to halo samples whose average mass is $\sim 1.3$ and
$5\times 10^{13}\hmsun$, respectively. The dotted and dashed curves
are formulae based on the excursion set theory, whereas  the solid
curves are fits motivated by the peak model.
\label{fig:bias}}
\end{figure}

\section{The large-scale bias of dark matter haloes}

The large-scale bias contains important information on the abundance
and clustering of biased tracers of the density field.  To compare
theoretical expectations with measurements of dark matter halo bias, I
will assume that peaks of height $\nu=\dc/\sigma_0(R,z)$ identified in
the initial, smoothed density field $\delta_M$ are associated with
objects of mass $M$ collapsing at redshift $z$.

The peak model predicts that, for moderate peak height, $b_\nu^{\rm
E}$ is significantly smaller than the value $1+\nu^2/\dc$ derived for
thresholded regions \cite{K84} due to  the correlation between the
peak height and the peak curvature \cite{BBKS}. However,  in the limit
$\nu\gg 1$, $b_\nu^{\rm E}(\nu)\approx 1+(\nu^2-3)/\dc$ which shows
that the evolved linear bias of initial density peaks of height $\nu$
indeed converges towards the prediction of \cite{K84}.  This should be
compared to well-known expressions derived from the extended
Press-Schechter formalism which, in the same limit, evaluate to
$b_{\rm MW}^{\rm E}(\nu)=1+(\nu^2-1)/\dc$ \cite{MW96}  and $b_{\rm
ST}^{\rm E}(\nu)\approx 1+(a\nu^2-1)/\dc$ \cite{ST99}. In the latter
case, $a=0.75$ follows from normalising the Sheth-Tormen mass function
to N-body simulations. Note that, whereas $b_{\rm MW}^{\rm E}$ and
$b_{\rm ST}^{\rm E}$ depend only upon the peak height, $b_\nu^{\rm E}$
is a function of both $\nu$ and $M$ (through $\gamma_1(M)$).

In Fig.~\ref{fig:bias}, these various predictions are compared with
measurements of the linear bias of massive haloes extracted from
numerical simulations of structure formation \cite{DSI09}. Error bars
show the scatter among various realisations. The measured halo bias
appears to depart from the Sheth-Tormen scaling at large $\nu$, in
agreement with recent measurements of the halo bias
\cite{CW08,T10}. Furthermore, the data shows evidence for a dependence
on $M$, but the exact magnitude of the effect is sensitive to the halo
finder. Because the best choice of filter is a matter of debate, I
treat $\gamma_1$ as a free parameter and show $b_\nu^{\rm
E}(\nu,\gamma_1)$ for $\gamma_1=0.4$ and 0.5 (a Gaussian filter yields
$\gamma_1\approx 0.65$ for the mass range considered), which provide a
reasonably good fit to the bias of $\bsim 2\sigma$ haloes. Note that
the peak expression $b_\nu^{\rm E}$ is also found to match the bias of
massive haloes in scale-free cosmologies rather well \cite{DWBS}.

\section{Peak biasing and the baryon acoustic oscillation}

Having checked that the peak model predicts a large-scale halo bias
$b_\nu^{\rm E}(z)$ consistent with simulations, I consider now the
impact of  the scale-dependent piece $b_\zeta^{\rm E}(z) k^2$. The
presence of such a term amplifies the contrast of the baryon acoustic
oscillation (BAO) in the correlation of initial density peaks relative
to that in the linear theory correlation \cite{D08}.
Eq.(\ref{eq:ppk}) can be used to estimate how much of this effect
survives  at virialization redshift (a more realistic calculation
should include the mode-coupling power).

\begin{figure}
\centering
\psfig{figure=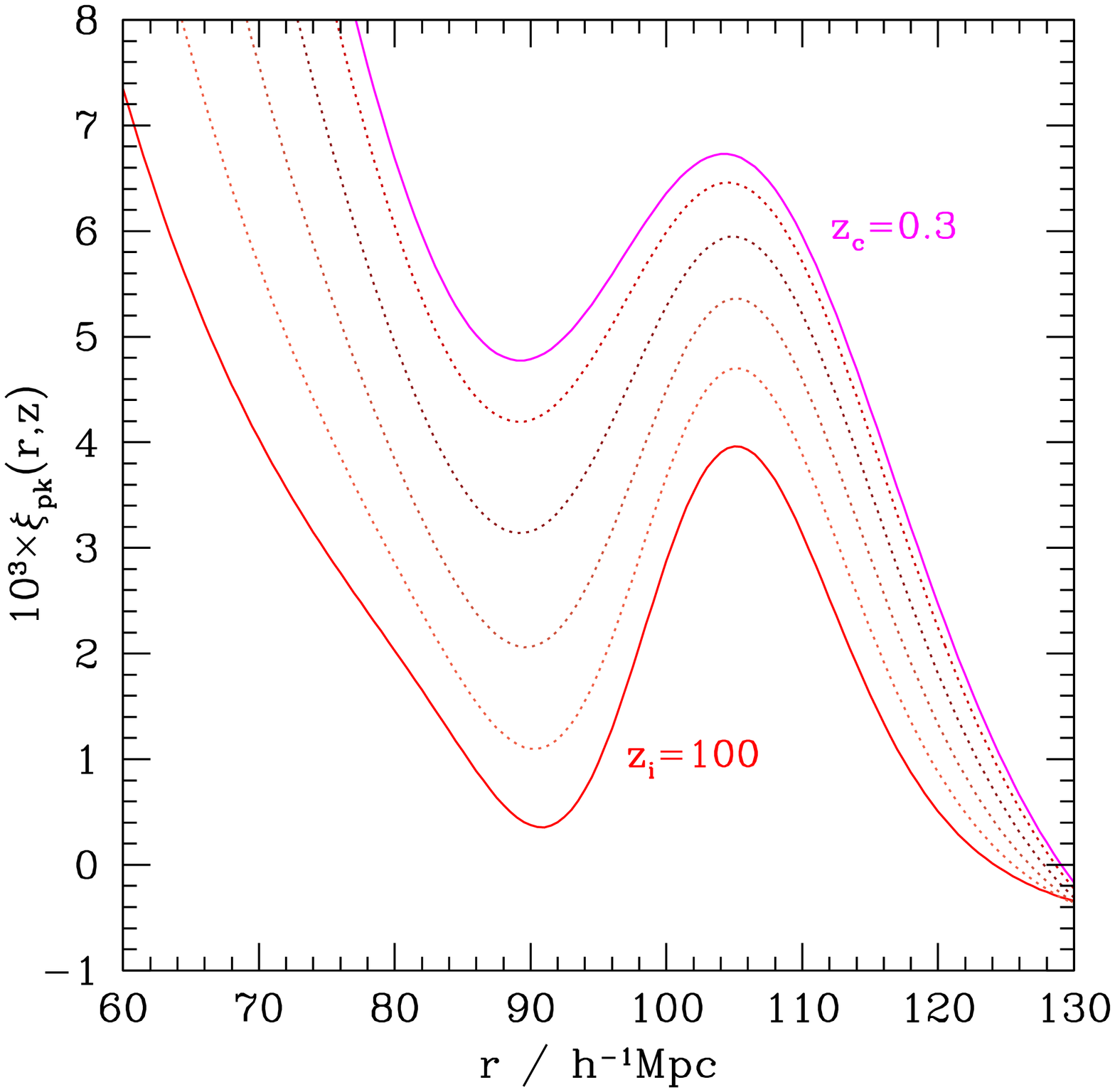,height=2in}
\psfig{figure=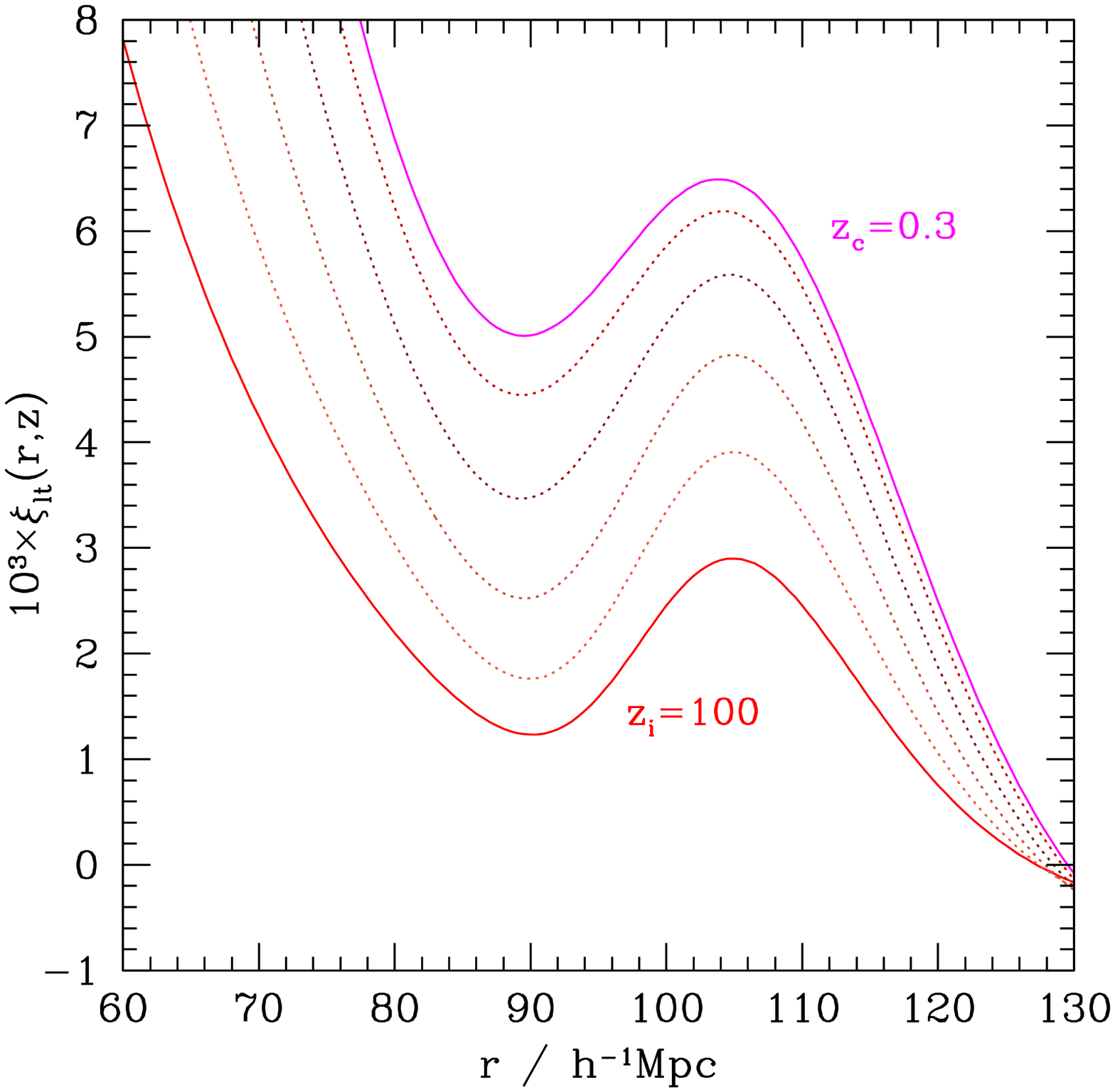,height=2in}
\psfig{figure=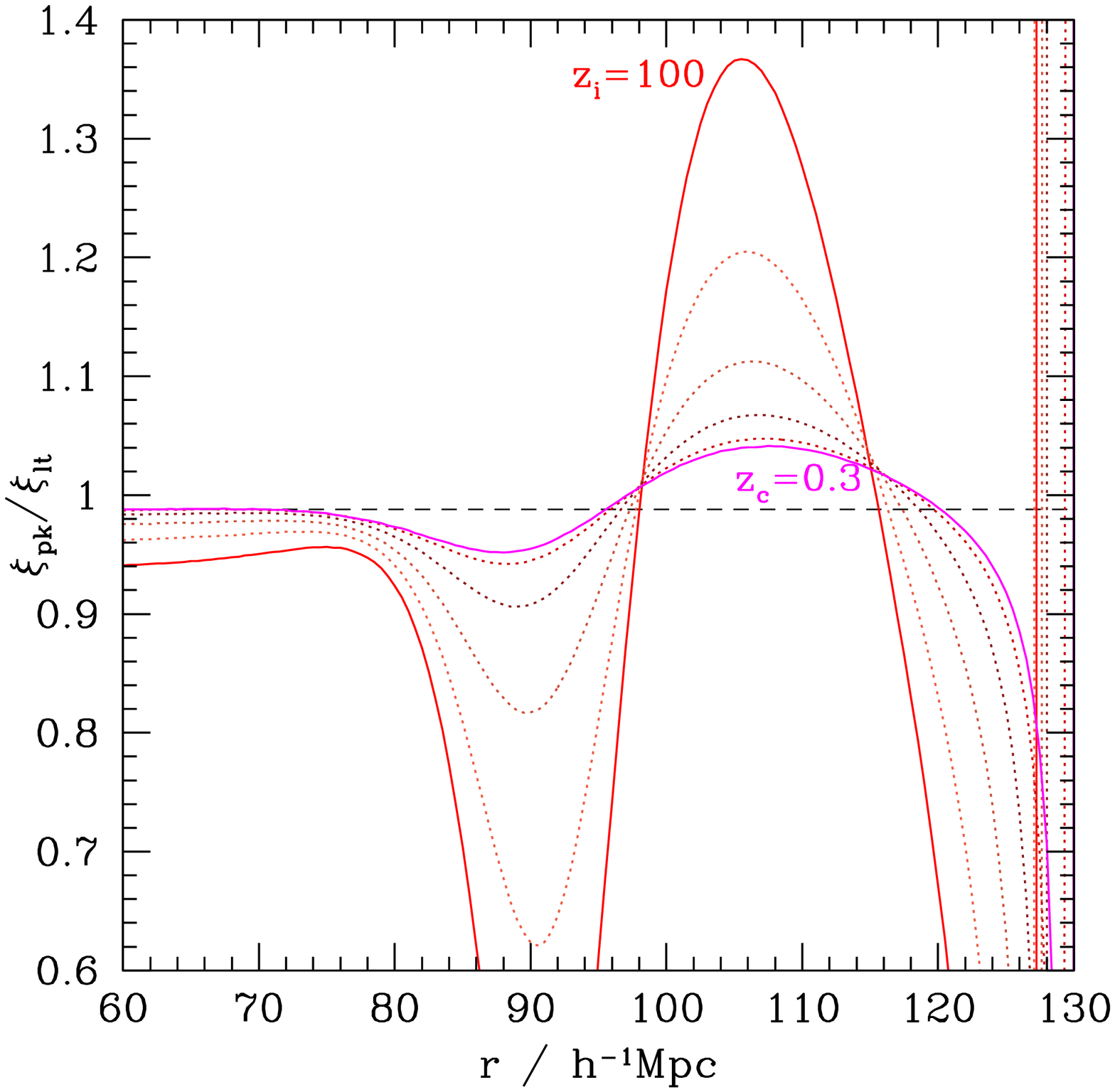,height=2in}
\caption{{\it Left}: Redshift evolution of the baryon acoustic
oscillation  in the correlation of initial, 2$\sigma$ density peaks as
predicted by  Eq.(\ref{eq:ppk}).  Results are shown at redshift
$z=100$,  5, 2, 1 0.5 and 0.3 (curves from  bottom to top). {\it
Middle}: Same as left panel but for ``linear tracers'',
for which the correlation simply is $b_\nu^{\rm E}(z)^2$ times the
evolved mass correlation {\it Right}: The ratio diverges at 
$r\sim 130\hmpc$ because zero-crossings do not coincide
\label{fig:bao}}
\end{figure}

To emphasise the effect of $b_\zeta^{\rm E}(z)k^2$, Fig.~\ref{fig:bao}
compares the redshift evolution of the large-scale, 2-point
correlation $\xi_{\rm pk}$ of initial density peaks (left) with that
of ``linear tracers'', $\xi_{\rm lt}$, for which $P_{\rm
lt}(k,z)\equiv G_\delta^2(k,z) [b_\nu^{\rm E}(z)]^2\,P_M(k,0)$
(middle).  The right panel displays the ratio between the two
correlations.  Results are shown for 2$\sigma$ density peaks
collapsing at $z_c=0.3$ and identified on a mass scale $5\times
10^{13}\hmsun$ with a  Gaussian filter. The relative amplification of
the BAO contrast in $\xpk(r,z_i)$ induces a scale-dependence in the
bias that decays with time owing to  the smearing from velocity
dispersion. At the collapse redshift however, the model predicts
residual scale-dependence across the BAO feature at the 5-10\% level
(right), a measurement of which in numerical simulations would provide
strong support for the validity of the peak approach.

\section*{Acknowledgments}

I would like to thank the organisers for a very enjoyable meeting.

\end{document}